\documentclass[12pt]{article}
\usepackage[utf8]{inputenc}
\usepackage{geometry}
\usepackage{amsmath}
\usepackage{graphicx}
\usepackage{array}
\graphicspath{ {./images/} }
\usepackage{biblatex}
\usepackage{setspace} 

\title{Sensorimotor Contingencies}
\author{Denizhan Pak}

\date{11 November 2023}

\begin{document}

\maketitle
\tableofcontents
    \section{Introduction and Motivation}
     4E views of cognition seek to replace many of the long-held assumptions of traditional cognitive science. One of the most radical shifts is the rejection of the sandwich model of cognition \cite{burr_embodied_2017}, which holds that mental processes are located between action and perception. Subversion of such a long-held assumption requires an accessible theoretical alternative with firm experimental support. One unifying thread among the emerging 4E camps is their shared insistence that sensorimotor contingencies (SMCs) are such an alternative.
     
	In the traditional view, representations of objects, scenes, and other agents are explicitly constructed from sensory information. These representations are the cognitive units upon which the algorithms of the brain act. This view implies representations should generalize across action-contexts and depend only on internal states of the brain and available sensory information. However, there are converging lines of evidence that negate such a passive theory of mental content by showing that perception, cognition, and action cannot be separated so neatly. This implies that the units of cognitive science should instead be integrated wholes of perception and action.

\begin{figure}
    \centering
    \includegraphics[width=1\linewidth]{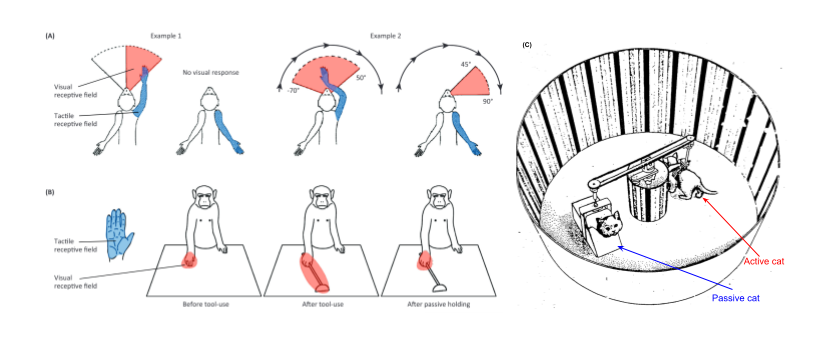}
    \caption{(A \& B) Dependence of receptive field on arm and tool positions. Adapted from \cite{engel_wheres_2013}. (C) Held-Hein experimental design. Only active cat gets visual input coupled to its own motions. Adapted from \cite{held_movement-produced_1963}.}
    \label{fig:enter-label}
\end{figure}
 
   The first piece of evidence for this differing viewpoint comes from measurements of neural activity. It is thought that neurons deeper in the visual stream have increasingly complex representations of perceptual information, eventually reaching the mythical grandmother cells. Although such concept cells have been found, their receptive properties have raised more questions than they have answered. A surprising and common finding is that the receptive fields of neurons across the brain change depending on action-context (Figures 1A-B) \cite{engel_wheres_2013}. For example, a neuron that responds to vertical edges at 40 degrees to the right of the visual field when the arm is centered may respond to vertical edges at 60 degrees when the arm reaches right. Several experiments have confirmed that many neurons that respond to visual information do not respond to specific portions of the visual field; they respond to positions in the visual field relative to the position of a limb. Experiments also show that positions of receptive fields can extend beyond the body. Neurons from a macaque cortex were shown to have receptive fields relative to the position of a tool the macaque was using. Thus, it appears that a neuron may have a tool-based receptive field and may even revert between an arm position- and tool position-based receptive field depending on whether a tool is actively being used \cite{stewart_enaction_2010}.

    Behavioral evidence also supports the view that action and perception are inseparably linked. The now famous Held-Hein experiments demonstrated that passive movements were insufficient for perceptual learning; active action was needed for cats to recognize edges in a visual cliff experiment (Figure 1C) \cite{held_movement-produced_1963}. Similar results have been shown in humans. In perceptual adaptation experiments, people require action-based feedback to adapt \cite{bingham_calibration_2014}. Sensory-substitution experiments where people have to learn a new sense-modality show the same conclusion \cite{bermejo_sensorimotor_2015}. It is not abstract perceptual information that allows organisms to learn but rather the interaction between action and perception. These results bring into question whether animals possess passive cognitive representations at all.
    
 Another motivating force comes from work on artificial agents which shows that representations of objects or scenes are not even required for many complex tasks: simpler perception/action-based solutions suffice. Tasks such as object categorization, referential communication, and navigation can all be achieved using perception/action loops coupled through neural dynamics \cite{beer_dynamical_2000}. A neural network in an action/perception loop can even drive a car using an onboard camera \cite{hasani_liquid_2020}. In cases where artificial agents are optimized to perform tasks, without additional model assumptions, optimal systems converge on action-specific representations of sensory information. These examples are further evidence that cognitive representations should not be assumed to be necessary for an agent performing intelligent behavior.
 
Although not yet conclusive, these lines of evidence are only part of the growing skepticism of the traditional cognitive view of minds as storing and computing abstract cognitive representations. Units of cognition must integrate action and perception and be able to account for the many complex behaviors of embodied agents. If we reject the traditional notion of passive cognitive representations, what is the unit of analysis for cognitive science? SMCs provide an answer.

    \section{What is an SMC?}
      An SMC is a quantifiable regularity between an action and its sensory consequences \cite{oregan_sensorimotor_2001}. The softness of a sponge comes from a person squeezing it in their hand. It could be quantified, if the exact amount of pressure applied and surface space covered by the fingers were known. But it is only through the action of the person that softness could be sensed.
 
	    SMCs have been studied by many researchers who did not have a term available to them to describe the way animals use actions to bring about specific sensations. This has led to conceptual confusion and difficulties in summarizing what is known about SMCs. Since the introduction of SMCs two decades ago, a significant effort has been put forth to formalize these ideas in computational and consistent terms.

     In their formalism, Buhrmann et al. \cite{buhrmann_dynamical_2013} identify two kinds of SMCs. All SMCs are relationships between an action and its sensory consequences. However, some SMCs are immediate results of an action and some are the result of an action that unfolds over time. This distinction matters because these concepts have different applications in experiments. An SMC which is the immediate consequence of an action as a result of physical laws is called a sensorimotor (SM) constraint. An SMC that is the result of an action over time is called an SM habit\footnote{Buhrman et al. refer to the set of all SM constraints which they call the SM environment and define the SM habitat as the set of all SM habits. They also identify two other uses of the term SMCs: SM coordinations (normative SM habits) and SM strategies (sets of SM coordinations). Since their original paper, usage of these terms has shifted and the current paper attempts to unite their original terminology with contemporary developments.}.

\begin{figure}
    \centering
    \includegraphics[width=0.8\linewidth]{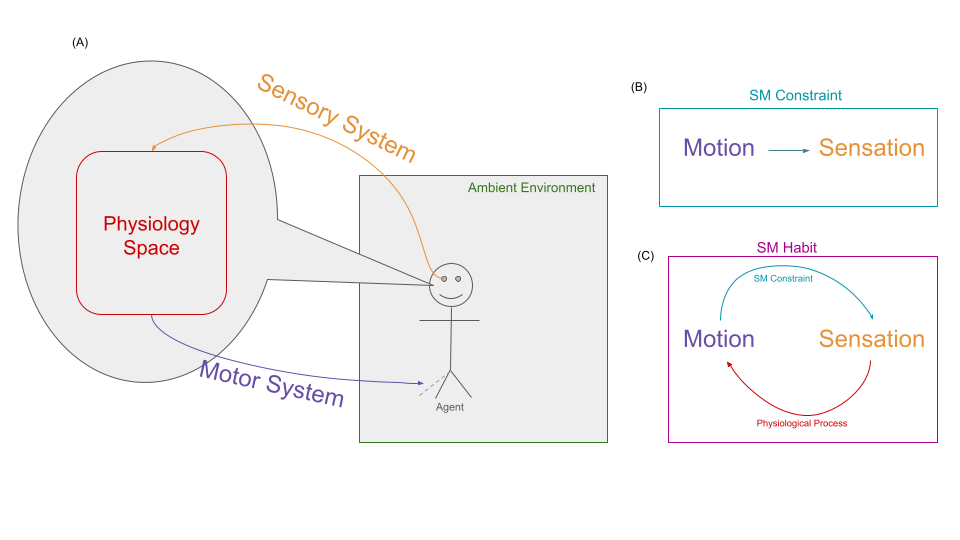}
    \caption{(A) A SM Environment is an ambient environment that contains an agent who has a sensory system which sends information to the internal physiology which directs the motor system which moves the agent. (B) A SM constraint is an SMC that does not incorporate the agent's own dynamics. (C) A SM Habit is an SMC that occurs over the course of an action/perception loop.}
    \label{fig:enter-label}
\end{figure}
 
	In order to describe the SMCs of an experimental system, we must identify several components.  An agent occupies an SM environment (Figure 2A). The agent is made up of a physiology space: its brain and body, a sensory system which carries information from the environment to its physiology, and a motor system that is driven by the physiology and causes the agent to move in the environment.

The SM environment contains SM constraints which are laws of how the movements of the agent will affect the agent's sensory system (Figure 2B). The agent performs SM habits which are actions over time that result in the agent reaching a specific sensory state (Figure 2A).

An example will help us show the difference between SM constraints and SM habits, which are related but not the same, and how to think about sensory systems, motor systems, and physiology spaces when looking at SMCs.

	The mobile paradigm, originally from the work of Carolyn Rovee-Collier \cite{kelso_coordination_2016}, can help elucidate what these terms mean. An infant's limb is connected by a ribbon to a mobile above. If the infant moves the limb, the mobile moves with it. The sensory system of the infant will measure the movements of the mobile through visual, auditory, and haptic feedback. The motor system will move the arm of the infant based on the output from the infant’s brain, i.e. physiology space. Every movement of the infant's arm has a physical effect on the mobile, this will cause a sensory input to the sensory system. The physical relationship between the movement of the arm and the resulting movement of the mobile is an SM constraint, and it is determined by the forces the limb applies to the ribbon and therefore the mobile; how much an arbitrary arm movement moves the mobile could be calculated using Newtonian physics. The infant, however, will not move her arm arbitrarily. She will develop a habit where she plays with the mobile, moving her arms up and down in her own pattern: a SM habit. 
 
    As scientists we wish to describe the mechanisms underlying the infant’s SM habit. To do this, we will need to explain how the physiology space interacts with the SM constraint through the sensory and motor systems to make-up the SM habit. This is why it is important to identify SM constraints as well as study the physiology space, sensory system, and motor system that make-up the agent.

In most real life contexts, we do not notice SM constraints. Move your head left, everything you see will move right, and that isn't surprising. The laws relating the motion and sensation are not mysterious. Optic flow is generated by bodily movement because of the laws of optics. But because they are so mundane, most SM constraints are not obvious. Scientific investigation is needed to identify them and to determine whether they are relevant to the system of interest. In practical terms, as we will see in section 3, SM constraints are an important starting point for 4E explanations of behavior.

Before moving on from terminology, let us consider one more example: catching a fly ball. A ball is in the air, and the agent intends to catch it. One strategy is to move so that the ball is translating up at a constant velocity and its visual size is growing at a constant rate \cite{sugar_unified_2006}. If the catcher maintains these sensory states, eventually the ball will land in her outstretched hand. There is a regular and quantifiable relationship between the action of the organism (following the ball) and the resulting sensory consequence (catching the ball). Catching a fly ball is an SM habit.

In this example, we can see how an SM constraint relates to an SM habit. The catcher must move left if she sees the ball veering in that direction. Moving left causes the visual image of the ball to move right which is the result of an SM constraint. By moving her body according to the appropriate SM constraint, the catcher maintains the necessary motion until the SM habit is complete. The agent uses her sensory system to measure the position of the ball. This measurement affects the agent's physiology which causes the agent to move in the appropriate direction; the end result of this process is the completion of the SM habit by sensing the ball being caught.

	This action requires that the agent’s physiology be attuned to the SM constraint of optic translation. Somehow the physiology of the agent “knows” that the sensory input of the ball being too far to the left, calls for moving left. We will discuss this attunement more in section 6.
 
	We can think about these same concepts in more computational terms using the formal theory of SMCs. In the case of the linear theory of SMCs \cite{angulo_dynamical_nodate}, we can consider the agent as possessing a vector that represents its current sensory state. At every time-step the agent chooses an action which changes the sensory state to another sensory state. This action is represented by a matrix. The physiology space chooses which action matrix comes next, using the current sensory state. The way an action matrix multiplies a sensory vector to generate a new sensory vector is an SM constraint and the sequence of vector and matrix multiplications over time is the SM habit.
 
	There are two important qualifications for this quantitative scheme. First, we currently assume that the only thing that changes a sensory state is the action of the agent; work on introducing a dynamic environment into this framework is an open research area. Second, real sensory and motor systems are non-linear. Conceptually, we can think about SMCs in terms of a sensory vector that is multiplied by an action matrix; for actually calculating the relevant quantities, though, a more general non-linear theory is needed. The beginnings of a non-linear theory of SMCs has been outlined in the thesis of J.L. Gordon \cite{godon_structuralist_2022} and is also an active research area.

\begin{table}[]
\begin{tabular}{|l|l|}
\hline
Sensoriotor Contingencies (SMCs) & \begin{tabular}[c]{@{}l@{}}Quantifiable regularities in an action and its\\ sensory consequences.\end{tabular}                                                                                                  \\ \hline
Sensorimotor Constraints         & \begin{tabular}[c]{@{}l@{}}SMCs where the action is a literal physical \\ movement and its consequences are \\ physical laws - downward optic flow from \\ looking up.\end{tabular}                             \\ \hline
Sensorimotor Habits              & \begin{tabular}[c]{@{}l@{}}SMCs where the action is extended through \\ time and involves an agent operating in an \\ action/perception loop - tracking a ball by \\ keeping it visually centered.\end{tabular} \\ \hline
Sensorimotor Environment         & \begin{tabular}[c]{@{}l@{}}The environment in which an agent can \\ perform an action/perception loop - made \\ up of many sensorimotor constraints.\end{tabular}                                               \\ \hline
Sensorimotor Habitat             & \begin{tabular}[c]{@{}l@{}}The lifelong trajectory of an agent through \\ the sensorimotor environment - made up \\ of many sensorimotor habits.\end{tabular}                                                   \\ \hline
Sensory Volume                   & \begin{tabular}[c]{@{}l@{}}The way in which an agent's sensory system\\ is embedded in the sensorimotor environment.\end{tabular}                                                                               \\ \hline
Attunement                   & \begin{tabular}[c]{@{}l@{}}The way in which an agent can use a sensorimotor\\ constraint to enact a sensorimotor habit.\end{tabular}                                                                               \\ \hline
\end{tabular}
\caption{SMC terminology}
\label{table:1}
\end{table}

The rest of this paper will explore the theory of SMCs across various contexts and demonstrate how it contributes to our scientific understanding of cognition.
In section 3, we will consider an application of the SMC framework and highlight how the different kinds of SMCs can help us make sense of behavior.
In section 4, we will look at SM environments, where SM constraints live, to get a better understanding of the function of SM constraints.
In section 5, we will consider the lifelong behavior of the organism and how this relates to its SM habits.
In section 6, we will look at how the relationship between an SM constraint and an SM habit is developed.
In section 7, we will turn to the history and philosophical implications of SMCs.
In section 8, we will discuss conclusions and highlight future directions.
    
    \section{SMCs in Action}
 SMCs can help us build a mechanistic understanding of cognition. The neuroethology of bees provides an example of how. Bees perform a range of complex tasks, from identifying flowers to navigating across forests. After a long day of foraging, a bee must enter through the opening of their hive. This is no easy feat: building drones capable of flying through narrow gaps using on-board cameras is a hard problem. We can turn to the work of Mandyam Srinivasan to see how bees solve this task \cite{srinivasan_how_1997}.
 
	First, it seems reasonable that bees would use vision to guide their flight. To test this, Srinivasan and colleagues designed a corridor where they placed a series of vertical markers along the right and left sides. They found that by manipulating the interval between markers, they were able to control how close the bee flew to one side of the corridor or the other. This discovery identifies the SM environment in which the bees operate. The bees use their visual sensory system to guide their flight. The velocity of the bee is determined by the motor system. The combination of a visual sensory space and velocity is a well-known SM environment: it is optic flow. 
 
  Optic flow also comes with SM constraints. Bees flying through an opening will experience an expansion of the edges of that opening in their visual field (as humans do when they walk through doors, and the left and right side of the door grow in the visual field); they will also experience equal amounts of light contrast hitting both their eyes. Though Srinivasan did not phrase his experimental goal as such, his next step was to figure out which of the SM constraints the bees were actively using. 
 
	Srinivasan and colleagues investigated which of the two SM constraints the bees use to enter the hive by fixing the color saturation of the markers to the size of the intervals, thus controlling for changes in overall contrast. If the bee's trajectory was affected by interval distance, then they would falsify the contrast-based hypothesis. They found that bees rely on the interval between markers and not contrast. The mechanism underlying the SM habit is based on controlling the expansion of the edges.
 
	Finally, Srinivasan and collaborators wanted to understand the physiology behind the bee's movement. They developed a motion detector circuit based on known neuroanatomy that predicted that early visual neurons producing phasic responses should saturate based on low contrast. Recording from the brain of an immobilized bee, this was exactly what they found. They were able to create a neurally mechanistic model of the physiology that drives the SM habit. Having described what is effectively the physiology space, they had identified the sensory and motor systems, the physiology space, and the SM constraints that apply to the SM habit: they had shown how SMCs allow us to scientifically and mathematically model the bee’s cognition. 
 
    	These results show a general pattern for how SMCs can be investigated. The first step is to identify an SM habit (in this case entering the hive). The second step is to identify the SM environment in which this habit takes place (in this case the optic flow). The third step was to identify which of the possible constraints in this SM environment the organism is using (in this case edge expansion). The fourth step is to propose and test a mechanistic model of how the organism’s physiology space interacts with the identified SM constraint (in this case how the bee’s brain measures the expansion of the edges).
     
	This four step process is the basis for how SMCs can contribute to our understanding of cognition. SMCs provide a unit for 4E cognitive science which can be studied in mechanistic detail using a theory/experiment loop. This sequence is not unique to the work on bee navigation; similar success stories can be found in many other species \cite{warren_dynamics_2006}. Although this is a standard approach and highlights the basic structure of SMC-based explanations, it is only one of the ways in which the concept of SMCs has contributed to our understanding of cognition.
    
    \section{SM Environments as Constraints}
    The constraints of an SM environment are not specific to any animal. They apply to any agent that measures the same sensory information and moves in similar ways. They can be useful for describing the SM habit of a given animal but also apply to comparative analysis across different animals. 
    
A recent example comes from the work of Dorris Tsao and collaborators on the question of object segmentation, a problem known to be hard for artificial vision systems \cite{tsao_topological_2022}. They demonstrate how an embodied agent in the optic flow SM environment can exploit the constraint of edge discontinuity between surfaces to solve the problem of segmentation. They provide a novel segmentation algorithm based on this SM constraint. In a Tsao and experimental collaborators later showed that humans, macaques, and lemurs use the identified strategy, while treeshrews and rats use a different texture-based strategy for object segmentation. The implication of these results is that different animals that occupy the same SM environment may use different constraints within that SM environment to navigate through it. Studying SM environments not only informs us about different animals but also about different evolutionary lineages.

	Rodents and primates are not limited to optic flow as their only SM environment. All animals possess multiple sensory and motor systems and simultaneously occupy several SM environments, each defined by a different pair of sensors and motors. One could apply these principles to any sensory and motor system, as long as a "vector" representation of the sensory space and a "matrix" representation of the motor space are available. 
 
	Given that there are so many possible SM environments and since the constraints of SM environments apply so broadly, it is tempting to ask if there are constraints that apply to multiple SM environments, constraints that we could apply to smell as well as vision. Such constraints could affect any agent, in any SM environment, from robots to animals to aliens. These constraints exist and have been incorporated into the mathematical theory of SMCs but require us to refine our understanding of the definitions from earlier.
 
We must first develop an intuition for how to visualize an SM environment. In the case of optic flow, the sensory space is the optic array, this is often visualized as a plane in front of the visual agent. This plane is embedded in the real three-dimensional world. The observer turns their head and rotates the plane. The action matrices change the sensory state of the organism by physically moving the position of the sensory manifold in ambient space\footnote{The term ambient space is used because it applies to both simulated and physically embodied agents. For physical agents, ambient space is the physical world.
}. 

\begin{figure}
    \centering
    \includegraphics[width=1\linewidth]{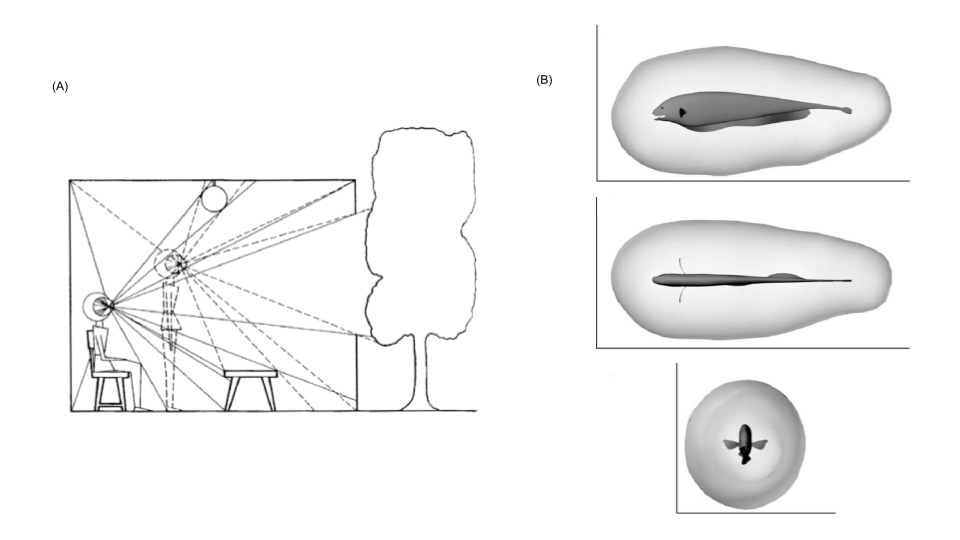}
    \caption{(A) The optic arrays of two sitting and standing observers. Adapted from \cite{tsao_topological_2022} (B) The electric sensory volume of an electric fish. Adapted from \cite{snyder_omnidirectional_2007}.}
    \label{fig:enter-label}
\end{figure}

For  a more explicit visualization, we can turn to a very different sensory system: the electrical sense of the electric fish. The fish produces an electrical field around its body, this allows it to sense conductive surfaces nearby. This field has been quantified as a cylinder in ambient space around the body of the fish \cite{snyder_omnidirectional_2007} (Figure 3B); movements of the body move the position of the sensory cylinder as a result.

 The visual plane is not actually the optic array, it is a cross section of the optic array. The optic array (Figure 3A)  could be visualized as a pair of cones shooting out of the eyes of the agent as far as visual acuity will allow. This three-dimensional volume is the true sensory space of optic flow. The moving observer moves their optic array. In general, we can refer to the subset of ambient space spanned by a sensory system as a sensory volume. The sensory system operates by taking a measurement of a portion of the ambient space that is spanned by the sensory volume. A person sees the surfaces that reflect light inside their visual cones. 
 
 We can refine our understanding of the motor system based on this way of viewing the sensory system. The motor system alters the position of the body in ambient space. This also moves the sensory volume in ambient space. 

	The concept of posture has a straightforward interpretation in this framework. It is the orientation of an agent's sensory volume in ambient space. Depending on the orientation, the sensory volume samples a different part of ambient space. The ambient environment with its inclines, obstacles, and physical laws constrain the set of postures an animal can take and therefore also constrain the positions a sensory volume can occupy.
 
    	Not only does the SM environment constrain posture but also movement. When an opening is below a critical size relative to a person's shoulder width, the person will rotate so as to keep one shoulder in front while moving through \cite{warren_dynamics_2006}. This results in an atypical egocentric viewpoint which alters the visual and auditory volume orientations relative to normal forward movement. The physical environment forces a postural configuration, which constrains the possible movements through the SM environment.
     
	Some researchers have considered how much information about ambient space can be derived purely from constraints on the movements and postures of a sensory volume. Mathematically it has been shown that a range of obstacles, the size of a room, and even the outline of an object can be determined without reference to sensory content and purely through interactions of the sensory volume \cite{godon_structuralist_2022}. Experimentally, participants were placed in a VR headset where the visual field flashed every time they looked at an invisible object. Purely from this discrete event feedback and head movements, participants were able to identify the rough contours of the object that they could not see directly \cite{suzuki_sensorimotor_2019}, this is another example of how constraints on visual volume can be used by an agent.
 
	We will consider one last example which demonstrates the breadth of this framework. The sensory volume is a volume in ambient space and, like any volume, we can calculate its size. If a sensory volume is small, then the agent must react to information in that sensory volume immediately, if it is large, the agent has time to plan. If something touches me, the time I have to make a decision is very short. If I see a lion on the horizon, I have plenty of time to pick my actions. The sensory volume is directly correlated with the decision-time available for information from that sensory system. Therefore, we should expect that sensory systems with large sensory volumes will be processed by larger networks with many local and recurrent feedback connections because there is more time available for the information to be integrated. This constraint applies to any sensory system of any agent.
 
	The sensory volume of vision underwater is significantly smaller than the sensory volume of vision on land due to the diffraction of light, light travels farther in air. If the size of their sensory volume truly constrains all agents, we should expect that it also constrains evolution itself. A prediction of this theory is that the expansion of the visual cortex should have happened before vertebrates arrived on land and after their eyes moved to the top of their heads, since the position of the eyes allowed them to see above water. This is exactly what Maciver and Finley observed in the evolutionary record \cite{maciver_neuroecology_2022}, the growth in visual cortex size happened before vertebrates ended up on land but after the increase in their visual volume.
 
	The constraints of the SM environment apply to all organisms that occupy it, although organisms may or may not actually make use of various constraints. The SM environment is part of the physical environment and also includes constraints that apply because of this physicality. The utility of studying the SM environment is that it helps us identify these constraints, and it can be used not only to predict the specific behaviors of an agent but also to predict the possible behaviors of all possible agents.
    
    \section{Principles of the SM Habitat}
    In our analysis of SMCs so far we have yet to truly consider the agent itself. In sections 2 and 3, we looked at specific SM habits and analyzed them as we would traditional cognitive tasks. In section 4, we focused on the structure of the SM environment. However, in 4E approaches, the agent is not simply a task-completion system, it is living and embodied. To appreciate this fact, we will define an additional concept: the SM habitat. The SM habitat is the lifelong trajectory of the agent through the SM environment. By considering the structure of the SM habitat we will see a surprising fact: introducing the high-dimensional and complex lifetime of the organism may initially seem to complicate our understanding of behavior, but, as it turns out, there are simpler and more fundamental principles that can only become apparent when we look at the complete agent-environment system over its lifetime.

The structure of the SM habitat is determined both by the physiology of the agent and the SM environment. The SM environment determines the possible postures and movements, while the physiology determines which of those possibilities are enacted by the motor system. This means that the dimensions of the SM habitat are the dimensions of the physiology space and SM environment combined. A complete characterization of the SM habitat of any organism has not yet been or potentially cannot even be accomplished.
Even with limited access, a combination of clever experimental design and creative theoretical exploration has already shed light on some of the structures that make up SM habitats. Three core principles provide a useful starting point:

\begin{enumerate}
    \item The SM habitat has a smaller set of core dimensions.
    \item SM habits are the syllables of the SM habitat.
    \item Cascades are the means of structural change of the SM habitat.
\end{enumerate}

We consider each in more detail.

The first principle is that the SM habitat is lower dimensional than it may initially seem. Some physiological variables are uninformative about how the sensory volume moves through ambient space; for instance, the dimensions specified by an individual ion channel may play a very insignificant role in how the whole body moves. More surprising is that among significant variables there is major redundancy. Dimensionality reduction techniques applied to behavioral and physiological measurement have shown that most dimensions of the SM habitat are highly correlated.

The movements of an animal can be predicted using only a handful of abstract dimensions, called eigenbehaviors \cite{eagle_eigenbehaviors_2009}. Work on c. elegans shows that the gaits of the worm vary along 4 axes that can be derived using dimensionality reduction techniques applied to behavioral recordings \cite{stephens_dimensionality_2008}. The existence of the eigenbehaviors make the SM habitat easier to analyze and hints that there is more structure to the SM habitat than meets the eye.

	A more intuitive example comes from brain physiology, where studying eigenmodes has become standard practice \cite{eagle_eigenbehaviors_2009}. Identifying the state of the brain does not require measuring every neuron and every ion channel. There are several core brain networks that can be thought of as being more or less active. By plotting the activity of each network as a dimension, the state of the brain can be specified using a low dimensional description. These dimensions are called eigenmodes. As in behavior, the existence of eigenmodes implies that brain activity is better predicted by this lower-dimensional space. 
 
	The existence of eigenmodes in behavior and physiology demonstrates the utility of studying the SM habitat. If we were to consider either behavior or physiology in isolation, we might have ignored this shared attribute. The lower dimensionality of either behavior or physiology is useful for their study, but the fact that both share this feature hints that there might be a more fundamentally shared cause.
 
	The second principle is about how the SM habitat is structured through time. We might imagine that the SM habitat is all over the place. Physiology is very complicated and even if the SM habitat has eigenmodes, we have no reason to expect that within this low-dimensional space there is some sort of pattern. However, it turns out that there is a common pattern to animal behavior.
 
	A near universal principle is that animal behavior is structured by a sequence of semi-fixed smaller action patterns \cite{datta_computational_2019}. we have been calling these motions SM habits. SM habits are abundant in animal behavior across many time-scales. Again, consider the bee entering its hive. Bees do not freely explore the world, occasionally entering their hives. They leave their hive, search an area, select and pollinate a flower, find their home, and only then enter their hive. It is by finding their home and seeing the hive that they position themselves to perform the task of entering it. Animals have goals like building a web, hunting for food, or getting home and they accomplish these goals by sequencing smaller units of (meta)stable behaviors, we call SM habits. Such a structure has been interpreted in terms of a habit network and has been a growing theme in animal behavior research \cite{datta_computational_2019}.
 
    This syllabic structure is also found in the physiology that guides behavior. Recordings from cortex show that the brain navigates through an energy landscape where it rolls into individual attractor basins which it occupies for a period of time, then gets pushed out only to fall into the next \cite{shine_adaptively_2022}. These attractor basins have been found to correspond to specific sensorimotor habits \cite{graziano_ethological_2016}. The brain can even be pushed into specific attractor basins that in turn result in the execution of a specific habit.

	SM habitats are built using a core set of syllables. These syllables are shared across both the physiology and behavior; they can be used to predict and even control an organism's movements. Identifying the network that makes up these syllables and the mechanisms of transition between them is an exciting front for future research in SMCs. If syllabic structure is truly a universal of adaptive behavior, theoretical work could explore the physical laws through which such structures may emerge.
 
	The third principle is that the SM habitat is structured by cascades. Cascades are landmark events in which the co-occurrence of various phenomena lead to significant long-term changes in the system overall. It seems to be the case that across different contexts cascades are a driving force for behaving systems and thus are fundamental to the SM habitat.
 
We will look at an example of a cascade in an SM habitat, which also has implications for how infants develop. Head-camera and behavioral recordings in infants show that infants use many different strategies to keep their behavior low-dimensional and operating in eigenmodes \cite{smith_developing_2018}. Infants also behave in sequences, spending bouts with individual toys, dropping them, and moving on to the next. The SM habitat of human infants is very similar to any other animal, it is a network of low-dimensional SM habits. Coordination between SM habits and the naming of objects by others during play is fundamental to name learning. The co-occurrence of the care-giver's gaze, voice, and the SM habits of the infant are the basis of the development of joint attention. Landmark events and their cascading consequences are considered a touchstone of human cognitive development \cite{smith_dynamic_2007}. Cascades also occur in the domains of physiology and the brain, such as neuronal avalanches \cite{fosque_evidence_2021}. Although the content of cascades differ in different parts of the agent-environment system, they are abound all over. The theoretical structure of cascades can inform a foundational theory of adaptive behavior.

The SM habitat is the lifetime trajectory of an agent through the SM environment. It is constrained by the SM environment but is guided by ongoing physiological activity which interacts with the sensory and motor systems. The SM habitat is not as high-dimensional as it may seem and is dominated by lower dimensional eigenmodes. The SM habitat is made up of sequences of behavioral nodes in a habit network. The structure of the SM habitat is also fundamentally tied to the occurrence of cascade events across different interacting systems. These are clues that point us to a more foundational understanding of agent-environment systems that can be achieved through the continued development of SMC research.
    
    \section{Modeling Attunement}
   In several cases, we saw that an SM habit could be understood as the exploitation of an SM constraint. Many SM habits turn out to be describable in terms of an SM constraint that is being used by an agent to control a variable to achieve a desired sensory state \cite{warren_dynamics_2006}. To be able to make use of an SM constraint in this way, an agent must be attuned to it. On a philosophical level, this is a question about knowledge and its role in behavior. The term attunement is chosen as a way to describe the relationship between SM constraints and SM habits without invoking a specific philosophic interpretation.

	Developmental perspectives on SMCs have shed light on different parts of attunement. Behavioral embryology and developmental neuroethology have explored how the growth of the body and the 
 brain shapes the formation of the SM habitat \cite{lux_gottlieb_2013}. These results explicate how the interplay of ongoing genetic programs, sensory experience, and neural activity mutually interact to shape the development of various SM habits. Work in sensorimotor learning has looked at how agents contend with having their SM environments altered or how they adapt to being exposed to completely new SM environments \cite{jacquey_sensorimotor_2019}. This work highlights the constraints on attunement and has helped isolate some of the features that facilitate it. Microgenetic studies have looked at how changes in SM habits and environments can lead to developmental cascades, which makes clearer the different ways in which higher-level cognitive functions depend on the attunement to the SM environment \cite{smith_dynamic_2007}.

	All of these approaches demonstrate the ways in which the SMC concept is useful for our developmental understanding of cognition. A thorough review of even one of these fields would be outside the scope of this paper. Instead, this section will focus on the sufficient conditions for attunement to occur at all.
 
For some animals, it is possible to model physiology space directly in terms of actual physiological quantities; however, for most animals, including humans, physiology space is too large to be modeled explicitly. Instead, various camps working on SMCs have proposed different theoretical principles that can be used to describe physiology space abstractly. A minimal requirement for an abstract description of physiology space is that it must show attunement. We will consider different physiology space models that show attunement and how they differ in their implications for our understanding of physiology space.

	Before turning to computational models, let us consider how real animals attune to their SM environment. Results from many different studies on the development of SM habits show a similar pattern. Over the course of attunement, agents first demonstrate a period of high variability within the relevant SM environment, then, after a period of time, settle on a stable behavioral pattern that they improve over iterations. One example comes from \cite{bermejo_sensorimotor_2015}. Bermejo et al. use a sensory-substitution paradigm where participants are asked to determine the shape in front of them. They are blindfolded and given a device that converts RGB-camera inputs into auditory signals. Different participants adopt different strategies, but one finding is consistent: participants first explore the device by trying different movements, then settle on 2-3 typical movements which they alternate to accomplish the task. This example demonstrates a near universal developmental trend \cite{lux_gottlieb_2013}.

 For any theoretical model of physiology space to be a viable candidate, it should at least demonstrate attunement. In the future, through theory/experiment loops, it will be possible to refine candidate models by how well they match the more detailed experimental data that has already been collected on attunement. Here, we consider models that have been tested to show any attunement at all to get an overview of the theoretical landscape.

The most popular class of models for attunement is a variation of reinforcement learning (RL). In RL, agents can navigate through the SM environment and are rewarded for reaching certain sensory states. Unlike traditional RL which learns only actions and their corresponding reward values, SMC based RL learns the reward values for action-sensation pairings. Like other RL systems, the agents are designed to maximize their reward. Over iteration, certain SM habits are reinforced because they result in rewarding outcomes, a phenomena resembling attunement \cite{hay_behavior_2018}.

	A surprising finding from these models is that agents which solve many tasks with complex bodies will naturally develop a network like SM habitat structure \cite{jacquey_sensorimotor_2019}. This is because certain postural configurations will act as important hubs for transitions between different behaviors, resulting in certain action-sensation pairs being reinforced because many rewarding trajectories pass through them.

    Since this model is a theoretical model of physiology space, we should consider its implications for the interpretation of physiology space. The environment influences the physiology through sensations and rewards. The physiology generates actions as a result of a sensory input and the physiology restructures itself based on rewards. This is similar to our description of the SM habitat in the previous section, except that it also introduces the concept of a reward. In this case, rewards are distinct from other sensations, they affect the physiology space fundamentally differently. The theoretical justifications for this view are influenced by the reward prediction error hypothesis for the dopamine system \cite{caras_non-sensory_2022}.
 
	Others have questioned whether postulating two kinds of inputs from the environment, each with distinct physiological effects is necessary. Instead, they argue the reward process can be attributed to the physiology itself by incorporating intrinsic motivation systems \cite{aubret_survey_2019}. These models are similar to RL models, in that physiology is understood to be maximizing some abstract objective or reward function. They are different in that the reward is not something the agent gets from the SM environment, the reward is part of the physiological process itself and the only thing that is gotten from the environment is the sensory state. The reward is often represented as some function that compares a predicted sensory outcome of an action with the actual resulting sensory outcome of that action. The interpretation is that the physiology contains an internal model of the SM environment. Depending on the success of predictions from the internal model, the agent alters its behavior and physiology to minimize some form of prediction error.

These models show attunement to novel SM environments and develop appropriate SM habitats. Results also show that by introducing predictions at multiple time-scales, it is possible to develop modular behavioral units that can be recombined in different contexts \cite{yamashita_emergence_2008}. Additionally, it is possible to incorporate physiological variables into the internal model such as energy, and once incorporated this can cause the agent to find different parts of the environment rewarding depending on its internal states \cite{aubret_survey_2019}.

	In this case, the relationship between the physiological space and the SM environment is reduced to a motor and sensory system. This aligns with how we have defined the SM habitat. These models interpret the physiology space as a statistical model of the SM environment with the goal of minimizing its prediction error. There are several theoretical justifications for this view. The most prominent is the free energy principle which relies on a statistical mechanics-like interpretation of high-dimensional physiology space \cite{buckley_free_2017}.

 Researchers more aligned with cybernetics take issue with this statistical interpretation. Their argument is that although a system maintaining homeostasis may look as if it is minimizing prediction error, what organisms are actually doing is trying to keep their own internal equilibrium \cite{buhrmann_dynamical_2013}. Models in this paradigm define physiology as keeping the system within some viability boundary. The agent is a homeostatic system that is placed in a SM environment and allowed to behave until its behavior converges to a regular pattern. These agents often have some need, such as a hunger-level, that they try to maintain by navigating through their environment. The results from this line of work are surprising, not only because these systems are capable of attuning with no explicit reinforcement, but also because a system which is not explicitly optimizing, under the right environmental conditions develops optimal behavior.

 	One illustrative example of how a control system can learn optimal behavior without explicit optimization, comes from a model by Egbert and Barandiaran \cite{egbert_using_2022}. They develop an agent that tries to repeat the trajectories it has previously taken. The system is then placed in a world where it can bounce a diamond or ball against a wall. The diamond is unlikely to bounce back because of its geometry, but the ball returns after colliding with the wall. The agent learns to approach and bounce the ball and to avoid the diamond. This happens because the ball, even though it introduces variation, provides a predictable pattern of interaction through which the agent can maintain its homeostasis. Surprisingly the agent learns to bounce the ball with near perfect precision, and is even able to catch it if it wasn't bounced perfectly centered. This happens because bouncing the ball well is the most stable configuration.

  Even more exciting is that the idea behind this model has been applied directly in the developmental literature. In the previously mentioned mobile paradigm, experimenters introduce an artificial SM environment. How does the infant attune to the appropriate SM constraints to play with the mobile?
  
	A computational approach from Kelso and Fuchs uses the homeostasis idea by modeling the baby's arm as a self-maintaining oscillator \cite{kelso_coordination_2022}. The arm causes the mobile to move, which excites the nervous system of the baby, which in turn excites the arm movements. As a result of trying to keep arm movement frequency stable, even as the incoming stimulation of the mobile excites the infant's brain, the infant adapts by generating a form of coordinated movement with the mobile. The model predicts that there are only two different stable patterns of play that infants could perform based on the physics of the mobile. These predictions have since been tested and supported by experimental evidence \cite{kelso_coordination_2022}. This shows that theoretical models are useful for helping scientists make conceptual leaps but once developed can be extended to model real experiments.

This class of models is compatible with our definition of the SM habitat, the agent possesses a physiology space that interacts with the environment through a sensory and a motor system. The difference between this and intrinsic motivation models is that in the homeostatic model, physiology is not understood as trying to minimize an error relative to the SM environment but is trying to keep its internal structure within its own intrinsic constraints.

Attunement to the SM environment has been thoroughly studied in the developmental literature and many of its specific properties have been identified, although few models have tried to directly address the experimental data. One issue is that different computational models have fundamentally different interpretations of how the physiology should be understood relative to the SM environment. However, even a simple homeostatic process seems to be sufficient to observe the phenomena of attunement.
    
    \section{Philosophical Considerations}
   SMCs are not only useful for unifying theory with experiment and behavior with physiology but also for demonstrating the interplay between philosophy and science. The history of SMCs is deeply intertwined with the philosophy of mind. The term itself comes from philosophers trying to answer philosophical questions using the ideas and examples developed by scientists. The discussion on SMCs in both science and philosophy has major implications for our understanding of the very nature of knowledge itself.
    
Several camps in embodied cognitive science have adopted the concept of the SMC. Although they share the view that SMCs are a better alternative to traditional tasks and cognitive representations, they disagree about how we should understand the relationship between the SM environment and the SM habitat. This disagreement is most obvious in their views on attunement, but it is fundamentally about how knowledge should be interpreted in the context of SMCs.
In this section, I will use the history of the idea of the SMCs to contextualize this debate and conclude with the implications from the computational work in the previous section.

	O'Regan and Noe introduced the notion of SMCs in a Brain and Behavioral Sciences target article \cite{oregan_sensorimotor_2001} where they argued their sensorimotor contingency theory (SMCT) could be used to simplify the hard problem of consciousness. The hard problem is that there seems to be an unbridgeable gap between scientific explanations of experience and the phenomenology of those experiences. O'Regan and Noe address part of the hard problem: the problem of why different sensory experiences \textit{feel} different. They argue that different sensory modalities occupy distinct SM environments, and that the phenomenological experience of a sensation is not the result of the physiological measurement of sensory content but rather of changes in the SM habitat. For instance, their theory predicts a way to make seeing green feel like seeing red through the use of VR and eye-tracking, a claim that would later be tested and verified \cite{godon_structuralist_2022}. Their introduction of SMCs made it possible to shift the position of conscious experience from inside the organism into the interaction between the agent and the environment.
 
	A much more thorough discussion of their argument as well as an analysis of the responses made to the target article is possible. SMCT has become a popular topic in the philosophy of consciousness and remains a highly debated issue. Since this paper is concerned with SMCs in cognition and not consciousness, we will limit discussion of SMCT to the brief sketch above. For our purposes, O'Regan and Noe's primary contribution is introducing an alternative space for the mind, where it is not located inside the organism but in the space between the organism and its environment.

Ironically, while O'Regan and Noe developed the concept of the SMC and brought it into the study of consciousness, the precursors of the idea can be found in considerations of cognition. Two major thinkers, Donald MacKay and Gilbert Ryle, are considered by O'Regan and Noe to be the primary sources of inspiration for the idea of SMCs. MacKay, known for his explorations on mechanisms and meaning, was writing about the information available in perception and described the idea that the sensory consequences of an organism's actions contain redundant information: both the physiological processes that led to the action and the sensory content from that action convey the same information \cite{mackay_theoretical_1962}. He suggested that the organism could make use of this redundancy to coordinate its internal needs with the environment. The influence of his ideas can be seen in the homeostatic models of attunement, and his focus on sources of spatial information in perception finds a strong affinity with the notion of a sensory volume. Ryle is cited by O'Regan and Noe for his chapter on sensations in his book The Idea of Mind \cite{ryle_concept_2000} where he argues that recognizing a thimble is not due to some mental thimble representation, but rather due to an attunement to the way that the image of the thimble changes on the retina as the person moves around it. Not only can he be explicitly credited for the concept of attunement, but Ryle's description is a precursor to the notion of the SM environment. We can see the core idea emerging; for MacKay, the agent-environment interaction contains redundancies which can serve the agent; for Ryle, the knowledge of the thimble is not in the agent but is in the way the agent moves when the thimble is present. MacKay and Ryle saw the potential for a space between the organism and its environment, that O'Regan and Noe explicitly named.

After their BBS article, O'Regan and Noe both continued to develop SMCT. While Noe's works focused on the philosophical underpinnings, O'Regan pursued both experimental and modeling work to explore its implications. Half a decade later, SMCs made their way into broader discussions within 4E cognitive science \cite{engel_wheres_2013}. In this context, their role in conscious experience was de-emphasized and they were seen as a means for understanding cognition. Various formalisms were introduced to make sense of the concepts, and over the last decade, an integration with robotics led to the development of a nearly unified theory of SM environments \cite{godon_structuralist_2022}, although the theory of physiology space and the SM habitat is still up for dispute.

What is at stake, philosophically, is whether humans possess knowledge about the world. Normally, we would point to cognitive representations as examples of ways in which humans know about the world, but the developmental evidence suggests that cognitive knowledge is built upon lower-level SM habits. This implies that abstract knowledge about the world is not actually about the world, but is instead about the ways that we interact with and move through the world. If attunement is achieved without a representation of the world, then there is no fundamental basis upon which knowledge about the world is built, it is always between the agent and environment. The upshot is that knowledge, in the traditional sense, may not exist \cite{van_duijn_principles_2006}.

It is unlikely that any single scientific experiment could distinguish between these theses; they rely as much on interpretations of evidence as they do evidence itself. However, science can change the terms of the debate. The fact that it is possible for a homeostatic system to display attunement (without an internal representation) is a major hurdle to any view that would look to preserve traditional knowledge as true by definition. Although science may not explicitly distinguish between these views, it will become clearer which of these positions is most useful for explaining the results of scientific investigations and proposing novel falsifiable hypotheses.
    
    \section{Discussion}
   In this paper, I have outlined the concept of the SMC. I have focused on the computational interpretation of SMCs that has been thoroughly extended since it was originally proposed by Buhrman et al. \cite{buhrmann_dynamical_2013}, I have demonstrated how this interpretation has been explored using computational models as well as experimental approaches, and I have positioned this scientific understanding in its philosophical context. 
   
    It should be clear that SMCs provide many opportunities to improve our understanding of cognition. By serving as an alternative to cognitive representations, SMCs give us a new unit of explanation that shifts the emphasis from the disembodied computational mind to the physical organism that adapts to the constraints set forth by its environment. This does not prevent further inquiry into the cognitive but highlights the interesting twists and turns that one must take to get there.

	SMCs are a clear example of how philosophy can play a role in science. In the case of SMCs, work on the hard problem of consciousness led to the development of a concept that can be rigorously studied and tested in the lab based on quantitative theoretical hypotheses. Philosophy can provide concepts for scientists that can be explored through the theory and experiment loop, and the scientific results of this inquiry can warrant more thorough philosophical investigation.
 
	Much work remains to be done to improve our understanding of SMCs. SM environments have been characterized in some sense-modalities but not others and finding ways to study SM environments in their most general form requires the continued development of novel mathematical techniques. Vision related SMCs can be explored more thoroughly in the context of Virtual Reality (VR) which provides opportunities for the manipulation of SM environments and could give new insights into how organisms navigate and adapt to those environments. There are many opportunities for quantifying and characterizing SM environments that have been constructed in the developmental literature and inversely new experimental paradigms could be designed that take advantage of the theoretical tools available from the theory of SM environments. Work characterizing SM habitats in terms of its eigenmodes could be done for many different species and their sensory, motor, and physiological systems. Automated tools for analysis and identification of SM habits could be used to study humans as well as other animals for which the behavioral repertoire has not yet been completely characterized. Finally, many experiments have elucidated the exact mechanisms and constraints on attunement, this leaves plenty of opportunities for computational work to model these results and identify unifying principles. Controlled-rearing experiments that make use of real-time recording technologies and VR are especially well-suited to push our theoretical understanding of attunement to its limits \cite{pak_newborn_nodate}.
 
	In addition to all the experimental and modeling opportunities offered by SMCs, there is still much philosophical discussion to be had. Confusion persists where knowledge resides between the SM environment and physiology space. These debates need not exist in sterile isolation. Continued discussion between philosophers and scientists seems like fertile ground for expanding and furthering our current understanding. Cognitive science found its origins in the lofty goal of creating a unified field of mind science, a field where philosophy, psychology, artificial intelligence, and neuroscience could interact and generate novel perspectives complementing and challenging ideas within each. SMCs are an extension of this goal. They provide an intersectional unit that can be studied from different directions to clarify what it really is that we call mind.

\newpage
\printbibliography

@article{jacquey_sensorimotor_2019,
	title = {Sensorimotor {Contingencies} as a {Key} {Drive} of {Development}: {From} {Babies} to {Robots}},
	volume = {13},
	issn = {1662-5218},
	shorttitle = {Sensorimotor {Contingencies} as a {Key} {Drive} of {Development}},
	url = {https://www.frontiersin.org/article/10.3389/fnbot.2019.00098/full},
	doi = {10.3389/fnbot.2019.00098},
	language = {en},
	urldate = {2023-05-23},
	journal = {Frontiers in Neurorobotics},
	author = {Jacquey, Lisa and Baldassarre, Gianluca and Santucci, Vieri Giuliano and O’Regan, J. Kevin},
	month = dec,
	year = {2019},
	pages = {98},
}

@article{eagle_eigenbehaviors_2009,
	title = {Eigenbehaviors: identifying structure in routine},
	volume = {63},
	issn = {1432-0762},
	shorttitle = {Eigenbehaviors},
	url = {https://doi.org/10.1007/s00265-009-0739-0},
	doi = {10.1007/s00265-009-0739-0},
	language = {en},
	number = {7},
	urldate = {2023-05-22},
	journal = {Behavioral Ecology and Sociobiology},
	author = {Eagle, Nathan and Pentland, Alex Sandy},
	month = may,
	year = {2009},
	pages = {1057--1066},
}

@article{buhrmann_dynamical_2013,
	title = {A {Dynamical} {Systems} {Account} of {Sensorimotor} {Contingencies}},
	volume = {4},
	issn = {1664-1078},
	url = {https://www.frontiersin.org/articles/10.3389/fpsyg.2013.00285},
	urldate = {2023-05-22},
	journal = {Frontiers in Psychology},
	author = {Buhrmann, Thomas and Di Paolo, Ezequiel and Barandiaran, Xabier},
	year = {2013},
}

@article{beer_dynamical_2000,
	title = {Dynamical approaches to cognitive science},
	volume = {4},
	number = {3},
	journal = {Trends in cognitive sciences},
	author = {Beer, Randall D.},
	year = {2000},
	note = {Publisher: Elsevier},
	pages = {91--99},
}

@article{bingham_calibration_2014,
	title = {Calibration is both functional and anatomical},
	volume = {40},
	issn = {1939-1277},
	doi = {10.1037/a0033458},
	journal = {Journal of Experimental Psychology: Human Perception and Performance},
	author = {Bingham, Geoffrey P. and Pan, Jing S. and Mon-Williams, Mark A.},
	year = {2014},
	note = {Place: US
Publisher: American Psychological Association},
	pages = {61--70},
}

@misc{aubret_survey_2019,
	title = {A survey on intrinsic motivation in reinforcement learning},
	url = {http://arxiv.org/abs/1908.06976},
	doi = {10.48550/arXiv.1908.06976},
	urldate = {2023-05-09},
	publisher = {arXiv},
	author = {Aubret, Arthur and Matignon, Laetitia and Hassas, Salima},
	month = nov,
	year = {2019},
	note = {arXiv:1908.06976 [cs]},
}

@article{egbert_using_2022,
	title = {Using enactive robotics to think outside of the problem-solving box: {How} sensorimotor contingencies constrain the forms of emergent autononomous habits},
	volume = {16},
	issn = {1662-5218},
	shorttitle = {Using enactive robotics to think outside of the problem-solving box},
	url = {https://www.frontiersin.org/articles/10.3389/fnbot.2022.847054/full},
	doi = {10.3389/fnbot.2022.847054},
	language = {en},
	urldate = {2023-05-09},
	journal = {Frontiers in Neurorobotics},
	author = {Egbert, Matthew D. and Barandiaran, Xabier E.},
	month = dec,
	year = {2022},
	pages = {847054},
}

@book{ryle_concept_2000,
	address = {Chicago, IL},
	title = {The {Concept} of {Mind}},
	isbn = {978-0-226-73296-1},
	url = {https://press.uchicago.edu/ucp/books/book/chicago/C/bo3684918.html},
	language = {en},
	urldate = {2023-11-11},
	publisher = {University of Chicago Press},
	author = {Ryle, Gilbert},
	month = dec,
	year = {2000},
}

@article{srinivasan_how_1997,
	title = {How bees exploit optic flow: behavioural experiments and neural models},
	volume = {337},
	shorttitle = {How bees exploit optic flow},
	url = {https://royalsocietypublishing.org/doi/10.1098/rstb.1992.0103},
	doi = {10.1098/rstb.1992.0103},
	number = {1281},
	urldate = {2023-11-11},
	journal = {Philosophical Transactions of the Royal Society of London. Series B: Biological Sciences},
	author = {Srinivasan, Mandyam V. and Gregory, Richard Langton and Barlow, Horace Basil and Frisby, J. P. and Horridge, George Adrian and Jeeves, M. A.},
	month = jan,
	year = {1997},
	note = {Publisher: Royal Society},
	pages = {253--259},
}

@article{kelso_coordination_2016,
	title = {The coordination dynamics of mobile conjugate reinforcement},
	volume = {110},
	issn = {1432-0770},
	doi = {10.1007/s00422-015-0676-0},
	language = {eng},
	number = {1},
	journal = {Biological Cybernetics},
	author = {Kelso, J. A. Scott and Fuchs, Armin},
	month = feb,
	year = {2016},
	pmid = {26759265},
	pages = {41--53},
}

@article{burr_embodied_2017,
	title = {Embodied {Decisions} and the {Predictive} {Brain}},
	doi = {10.15502/9783958573086},
	author = {Burr, Christopher},
	month = mar,
	year = {2017},
}

@article{smith_developing_2018,
	title = {The {Developing} {Infant} {Creates} a {Curriculum} for {Statistical} {Learning}},
	volume = {22},
	doi = {10.1016/j.tics.2018.02.004},
	journal = {Trends in Cognitive Sciences},
	author = {Smith, Linda and Jayaraman, Swapnaa and Clerkin, Elizabeth and Yu, Chen},
	month = mar,
	year = {2018},
}

@article{held_movement-produced_1963,
	title = {Movement-produced stimulation in the development of visually guided behavior},
	volume = {56},
	issn = {0021-9940},
	doi = {10.1037/h0040546},
	number = {5},
	journal = {Journal of Comparative and Physiological Psychology},
	author = {Held, Richard and Hein, Alan},
	year = {1963},
	note = {Place: US
Publisher: American Psychological Association},
	pages = {872--876},
}

@article{angulo_dynamical_nodate,
	title = {On {Dynamical} {Systems} for {Sensorimotor} {Contingencies}. {A} {First} {Approach} from {Control} {Engineering}},
	language = {en},
	author = {Angulo, Cecilio and Acevedo-Valle, Juan M},
}

@article{tsao_topological_2022,
	title = {A topological solution to object segmentation and tracking},
	volume = {119},
	issn = {1091-6490},
	doi = {10.1073/pnas.2204248119},
	language = {eng},
	number = {41},
	journal = {Proceedings of the National Academy of Sciences of the United States of America},
	author = {Tsao, Thomas and Tsao, Doris Y.},
	month = oct,
	year = {2022},
	pmid = {36201537},
	pmcid = {PMC9564096},
	pages = {e2204248119},
}

@article{bermejo_sensorimotor_2015,
	title = {Sensorimotor strategies for recognizing geometrical shapes: a comparative study with different sensory substitution devices},
	volume = {6},
	issn = {1664-1078},
	shorttitle = {Sensorimotor strategies for recognizing geometrical shapes},
	url = {https://www.frontiersin.org/articles/10.3389/fpsyg.2015.00679},
	urldate = {2023-10-29},
	journal = {Frontiers in Psychology},
	author = {Bermejo, Fernando and Di Paolo, Ezequiel A. and Hüg, Mercedes X. and Arias, Claudia},
	year = {2015},
}

@article{sugar_unified_2006,
	title = {A unified fielder theory for interception of moving objects either above or below the horizon},
	volume = {13},
	issn = {1531-5320},
	url = {https://doi.org/10.3758/BF03194018},
	doi = {10.3758/BF03194018},
	language = {en},
	number = {5},
	urldate = {2023-10-15},
	journal = {Psychonomic Bulletin \& Review},
	author = {Sugar, Thomas G. and Mcbeath, Michael K. and Wang, Zheng},
	month = oct,
	year = {2006},
	pages = {908--917},
}

@article{oregan_sensorimotor_2001,
	title = {A {Sensorimotor} {Account} of {Vision} and {Visual} {Consciousness}},
	volume = {24},
	doi = {10.1017/s0140525x01000115},
	number = {5},
	journal = {Behavioral and Brain Sciences},
	author = {O?Regan, J. Kevin and Noë, Alva},
	year = {2001},
	pages = {883--917},
}

@article{hay_behavior_2018,
	title = {Behavior {Is} {Everything}: {Towards} {Representing} {Concepts} with {Sensorimotor} {Contingencies}},
	volume = {32},
	copyright = {Copyright (c)},
	issn = {2374-3468},
	shorttitle = {Behavior {Is} {Everything}},
	url = {https://ojs.aaai.org/index.php/AAAI/article/view/11547},
	doi = {10.1609/aaai.v32i1.11547},
	language = {en},
	number = {1},
	urldate = {2023-10-08},
	journal = {Proceedings of the AAAI Conference on Artificial Intelligence},
	author = {Hay, Nicholas and Stark, Michael and Schlegel, Alexander and Wendelken, Carter and Park, Dennis and Purdy, Eric and Silver, Tom and Phoenix, D. Scott and George, Dileep},
	month = apr,
	year = {2018},
	note = {Number: 1},
}

@article{engel_wheres_2013,
	title = {Where's the action? {The} pragmatic turn in cognitive science},
	volume = {17},
	issn = {1364-6613, 1879-307X},
	shorttitle = {Where's the action?},
	url = {https://www.cell.com/trends/cognitive-sciences/abstract/S1364-6613(13)00071-5},
	doi = {10.1016/j.tics.2013.03.006},
	language = {English},
	number = {5},
	urldate = {2023-10-08},
	journal = {Trends in Cognitive Sciences},
	author = {Engel, Andreas K. and Maye, Alexander and Kurthen, Martin and König, Peter},
	month = may,
	year = {2013},
	note = {Publisher: Elsevier},
	pages = {202--209},
}

@incollection{mackay_theoretical_1962,
	address = {Boston, MA},
	title = {Theoretical {Models} of {Space} {Perception}},
	isbn = {978-1-4899-6584-4},
	url = {https://doi.org/10.1007/978-1-4899-6584-4_5},
	language = {en},
	urldate = {2023-10-08},
	booktitle = {Aspects of the {Theory} of {Artificial} {Intelligence}: {The} {Proceedings} of the {First} {International} {Symposium} on {Biosimulation} {Locarno}, {June} 29 – {July} 5, 1960},
	publisher = {Springer US},
	author = {MacKay, Donald M.},
	editor = {Muses, C. A. and McCulloch, W. S.},
	year = {1962},
	doi = {10.1007/978-1-4899-6584-4_5},
	pages = {83--103},
}

@inproceedings{suzuki_sensorimotor_2019,
	title = {Sensorimotor contingency modulates visual awareness of virtual {3D} objects},
	url = {https://dx.doi.org/10.1162/isal_a_00143},
	doi = {10.1162/isal_a_00143},
	language = {en},
	urldate = {2023-10-08},
	publisher = {MIT Press},
	author = {Suzuki, Keisuke and Schwartzman, David J. and Augusto, Rafael and Seth, Anil K.},
	month = jul,
	year = {2019},
	pages = {70--71},
}

@phdthesis{godon_structuralist_2022,
	type = {phdthesis},
	title = {A {Structuralist} {Formal} {Account} of {Sensorimotor} {Contingencies} in {Perception}},
	url = {https://hal.science/tel-04000131},
	language = {en},
	urldate = {2023-10-08},
	school = {Sorbonne Université},
	author = {Godon, Jean-Merwan},
	month = jan,
	year = {2022},
}

@article{kelso_coordination_2022,
	title = {On the coordination dynamics of (animate) moving bodies},
	volume = {3},
	issn = {2632-072X},
	url = {https://iopscience.iop.org/article/10.1088/2632-072X/ac7caf},
	doi = {10.1088/2632-072X/ac7caf},
	language = {en},
	number = {3},
	urldate = {2023-10-05},
	journal = {Journal of Physics: Complexity},
	author = {Kelso, J A Scott},
	month = sep,
	year = {2022},
	pages = {031001},
}

@article{pak_newborn_nodate,
	title = {A newborn embodied {Turing} test for view-invariant object recognition},
	language = {en},
	author = {Pak, Denizhan and Lee, Donsuk and Wood, Samantha M W and Wood, Justin N},
}

@article{graziano_ethological_2016,
	title = {Ethological {Action} {Maps}: {A} {Paradigm} {Shift} for the {Motor} {Cortex}},
	volume = {20},
	issn = {1364-6613, 1879-307X},
	shorttitle = {Ethological {Action} {Maps}},
	url = {https://www.cell.com/trends/cognitive-sciences/abstract/S1364-6613(15)00274-0},
	doi = {10.1016/j.tics.2015.10.008},
	language = {English},
	number = {2},
	urldate = {2023-09-11},
	journal = {Trends in Cognitive Sciences},
	author = {Graziano, Michael S. A.},
	month = feb,
	year = {2016},
	pmid = {26628112},
	note = {Publisher: Elsevier},
	pages = {121--132},
}

@article{datta_computational_2019,
	title = {Computational {Neuroethology}: {A} {Call} to {Action}},
	volume = {104},
	issn = {08966273},
	shorttitle = {Computational {Neuroethology}},
	url = {https://linkinghub.elsevier.com/retrieve/pii/S0896627319308414},
	doi = {10.1016/j.neuron.2019.09.038},
	language = {en},
	number = {1},
	urldate = {2023-09-11},
	journal = {Neuron},
	author = {Datta, Sandeep Robert and Anderson, David J. and Branson, Kristin and Perona, Pietro and Leifer, Andrew},
	month = oct,
	year = {2019},
	pages = {11--24},
}

@article{lux_gottlieb_2013,
	title = {With {Gottlieb} beyond {Gottlieb}: {The} {Role} of {Epigenetics} in {Psychobiological} {Development}},
	volume = {7},
	issn = {2192001X},
	shorttitle = {With {Gottlieb} beyond {Gottlieb}},
	url = {https://www.medra.org/servlet/aliasResolver?alias=iospress&doi=10.3233/DEV-1300073},
	doi = {10.3233/DEV-1300073},
	language = {en},
	number = {2},
	urldate = {2023-08-24},
	journal = {International Journal of Developmental Science},
	author = {Lux, Vanessa},
	year = {2013},
	pages = {69--78},
}

@article{van_duijn_principles_2006,
	title = {Principles of {Minimal} {Cognition}: {Casting} {Cognition} as {Sensorimotor} {Coordination}},
	volume = {14},
	issn = {1059-7123},
	shorttitle = {Principles of {Minimal} {Cognition}},
	url = {https://doi.org/10.1177/105971230601400207},
	doi = {10.1177/105971230601400207},
	language = {en},
	number = {2},
	urldate = {2023-06-08},
	journal = {Adaptive Behavior},
	author = {van Duijn, Marc and Keijzer, Fred and Franken, Daan},
	month = jun,
	year = {2006},
	note = {Publisher: SAGE Publications Ltd STM},
	pages = {157--170},
}

@article{snyder_omnidirectional_2007,
	title = {Omnidirectional {Sensory} and {Motor} {Volumes} in {Electric} {Fish}},
	volume = {5},
	issn = {1545-7885},
	url = {https://journals.plos.org/plosbiology/article?id=10.1371/journal.pbio.0050301},
	doi = {10.1371/journal.pbio.0050301},
	language = {en},
	number = {11},
	urldate = {2023-05-02},
	journal = {PLOS Biology},
	author = {Snyder, James B. and Nelson, Mark E. and Burdick, Joel W. and MacIver, Malcolm A.},
	month = nov,
	year = {2007},
	note = {Publisher: Public Library of Science},
	pages = {e301},
}

@book{stewart_enaction_2010,
	title = {Enaction: {Toward} a {New} {Paradigm} for {Cognitive} {Science}},
	isbn = {978-0-262-01460-1},
	shorttitle = {Enaction},
	language = {en},
	publisher = {MIT Press},
	author = {Stewart, John and Gapenne, Olivier and Paolo, Ezequiel A. Di},
	year = {2010},
}

@article{warren_dynamics_2006,
	title = {The dynamics of perception and action.},
	volume = {113},
	issn = {1939-1471, 0033-295X},
	url = {http://doi.apa.org/getdoi.cfm?doi=10.1037/0033-295X.113.2.358},
	doi = {10.1037/0033-295X.113.2.358},
	language = {en},
	number = {2},
	urldate = {2023-04-28},
	journal = {Psychological Review},
	author = {Warren, William H.},
	year = {2006},
	pages = {358--389},
}

@article{caras_non-sensory_2022,
	title = {Non-sensory {Influences} on {Auditory} {Learning} and {Plasticity}},
	volume = {23},
	issn = {1438-7573},
	url = {https://doi.org/10.1007/s10162-022-00837-3},
	doi = {10.1007/s10162-022-00837-3},
	language = {en},
	number = {2},
	urldate = {2023-04-27},
	journal = {Journal of the Association for Research in Otolaryngology},
	author = {Caras, Melissa L. and Happel, Max F. K. and Chandrasekaran, Bharath and Ripollés, Pablo and Keesom, Sarah M. and Hurley, Laura M. and Remage-Healey, Luke and Holt, Lori L. and Wright, Beverly A.},
	month = apr,
	year = {2022},
	pages = {151--166},
}

@article{shine_adaptively_2022,
	title = {Adaptively navigating affordance landscapes: {How} interactions between the superior colliculus and thalamus coordinate complex, adaptive behaviour},
	volume = {143},
	issn = {0149-7634},
	shorttitle = {Adaptively navigating affordance landscapes},
	url = {https://www.sciencedirect.com/science/article/pii/S0149763422004109},
	doi = {10.1016/j.neubiorev.2022.104921},
	language = {en},
	urldate = {2023-04-26},
	journal = {Neuroscience \& Biobehavioral Reviews},
	author = {Shine, James M.},
	month = dec,
	year = {2022},
	pages = {104921},
}

@article{maciver_neuroecology_2022,
	title = {The neuroecology of the water-to-land transition and the evolution of the vertebrate brain},
	volume = {377},
	issn = {0962-8436, 1471-2970},
	url = {https://royalsocietypublishing.org/doi/10.1098/rstb.2020.0523},
	doi = {10.1098/rstb.2020.0523},
	language = {en},
	number = {1844},
	urldate = {2023-04-20},
	journal = {Philosophical Transactions of the Royal Society B: Biological Sciences},
	author = {MacIver, Malcolm A. and Finlay, Barbara L.},
	month = feb,
	year = {2022},
	pages = {20200523},
}

@article{fosque_evidence_2021,
	title = {Evidence for quasicritical brain dynamics},
	volume = {126},
	issn = {0031-9007, 1079-7114},
	url = {http://arxiv.org/abs/2010.02938},
	doi = {10.1103/PhysRevLett.126.098101},
	language = {en},
	number = {9},
	urldate = {2022-10-30},
	journal = {Physical Review Letters},
	author = {Fosque, Leandro and Williams-Garcia, Rashid V. and Beggs, John M. and Ortiz, Gerardo},
	month = mar,
	year = {2021},
	note = {arXiv:2010.02938 [cond-mat, physics:nlin, physics:physics]},
	pages = {098101},
}

@article{smith_dynamic_2007,
	title = {The dynamic lift of developmental process},
	volume = {10},
	issn = {1467-7687},
	url = {https://onlinelibrary.wiley.com/doi/abs/10.1111/j.1467-7687.2007.00565.x},
	doi = {10.1111/j.1467-7687.2007.00565.x},
	language = {en},
	number = {1},
	urldate = {2022-09-30},
	journal = {Developmental Science},
	author = {Smith, Linda B. and Breazeal, Cynthia},
	year = {2007},
	note = {\_eprint: https://onlinelibrary.wiley.com/doi/pdf/10.1111/j.1467-7687.2007.00565.x},
	pages = {61--68},
}

@article{hasani_liquid_2020,
	title = {Liquid {Time}-constant {Networks}},
	url = {http://arxiv.org/abs/2006.04439},
	language = {en},
	urldate = {2021-10-24},
	journal = {arXiv:2006.04439 [cs, stat]},
	author = {Hasani, Ramin and Lechner, Mathias and Amini, Alexander and Rus, Daniela and Grosu, Radu},
	month = dec,
	year = {2020},
	note = {arXiv: 2006.04439},
}

@article{stephens_dimensionality_2008,
	title = {Dimensionality and {Dynamics} in the {Behavior} of {C}. elegans},
	volume = {4},
	issn = {1553-7358},
	url = {https://journals.plos.org/ploscompbiol/article?id=10.1371/journal.pcbi.1000028},
	doi = {10.1371/journal.pcbi.1000028},
	language = {en},
	number = {4},
	urldate = {2021-10-19},
	journal = {PLOS Computational Biology},
	author = {Stephens, Greg J. and Johnson-Kerner, Bethany and Bialek, William and Ryu, William S.},
	month = apr,
	year = {2008},
	note = {Publisher: Public Library of Science},
	pages = {e1000028},
}

@article{yamashita_emergence_2008,
	title = {Emergence of {Functional} {Hierarchy} in a {Multiple} {Timescale} {Neural} {Network} {Model}: {A} {Humanoid} {Robot} {Experiment}},
	volume = {4},
	issn = {1553-7358},
	shorttitle = {Emergence of {Functional} {Hierarchy} in a {Multiple} {Timescale} {Neural} {Network} {Model}},
	url = {https://dx.plos.org/10.1371/journal.pcbi.1000220},
	doi = {10.1371/journal.pcbi.1000220},
	language = {en},
	number = {11},
	urldate = {2021-06-23},
	journal = {PLoS Computational Biology},
	author = {Yamashita, Yuichi and Tani, Jun},
	editor = {Sporns, Olaf},
	month = nov,
	year = {2008},
	pages = {e1000220},
}

@article{buckley_free_2017,
	title = {The free energy principle for action and perception: {A} mathematical review},
	shorttitle = {The free energy principle for action and perception},
	url = {http://arxiv.org/abs/1705.09156},
	urldate = {2020-09-30},
	journal = {arXiv:1705.09156 [q-bio]},
	author = {Buckley, Christopher L. and Kim, Chang Sub and McGregor, Simon and Seth, Anil K.},
	month = may,
	year = {2017},
	note = {arXiv: 1705.09156},
}
\end{document}